\documentclass[aps,prx, twocolumn, showpacs,superscriptaddress,floatfix]{revtex4-1}

\usepackage{amsmath,amssymb}
\usepackage{graphicx}
\usepackage{color}
\usepackage{rotating}
\usepackage{bm}
\usepackage{setspace}
\usepackage{hyperref}

\hypersetup{
colorlinks=true,
urlcolor= blue,
citecolor=blue,
linkcolor= blue,
bookmarks=true,
bookmarksopen=false,
}

\renewcommand{\vec}[1]{\bm{#1}}

\begin{document}


\title{Bounds on Nanoscale Nematicity in Single-Layer FeSe/SrTiO$_3$}

\author{Dennis Huang}
\affiliation{Department of Physics, Harvard University, Cambridge, Massachusetts 02138, USA}
\author{Tatiana A. Webb}
\affiliation{Department of Physics, Harvard University, Cambridge, Massachusetts 02138, USA}
\affiliation{Department of Physics \& Astronomy, University of British Columbia, Vancouver, British Columbia V6T 1Z1, Canada}
\author{Shiang Fang}
\affiliation{Department of Physics, Harvard University, Cambridge, Massachusetts 02138, USA}
\author{Can-Li Song}
\altaffiliation[]{Present address: State Key Laboratory of Low-Dimensional Quantum Physics, Department of Physics, Tsinghua University, Beijing 100084, China}
\affiliation{Department of Physics, Harvard University, Cambridge, Massachusetts 02138, USA}
\author{\\Cui-Zu Chang}
\affiliation{Francis Bitter Magnet Laboratory, Massachusetts Institute of Technology, Cambridge, Massachusetts 02139, USA}
\author{Jagadeesh S. Moodera}
\affiliation{Francis Bitter Magnet Laboratory, Massachusetts Institute of Technology, Cambridge, Massachusetts 02139, USA}
\affiliation{Department of Physics, Massachusetts Institute of Technology, Cambridge, Massachusetts 02139, USA}
\author{Efthimios Kaxiras}
\affiliation{Department of Physics, Harvard University, Cambridge, Massachusetts 02138, USA}
\affiliation{John A. Paulson School of Engineering and Applied Sciences, Harvard University, Cambridge, Massachusetts 02138, USA}
\author{Jennifer E. Hoffman}
\email[]{jhoffman@physics.ubc.ca}
\affiliation{Department of Physics, Harvard University, Cambridge, Massachusetts 02138, USA}
\affiliation{Department of Physics \& Astronomy, University of British Columbia, Vancouver, British Columbia V6T 1Z1, Canada}

\date{\today}


\begin{abstract}
We use scanning tunneling microscopy (STM) and quasiparticle interference (QPI) imaging to investigate the low-energy orbital texture of single-layer FeSe/SrTiO$_3$. We develop a $T$-matrix model of multi-orbital QPI to disentangle scattering intensities from Fe $3d_{xz}$ and $3d_{yz}$ bands, enabling the use of STM as a nanoscale detection tool of nematicity. By sampling multiple spatial regions of a single-layer FeSe/SrTiO$_3$ film, we quantitatively exclude static $xz/yz$ orbital ordering with domain size larger than $\delta r^2$ = 20 nm $\times$ 20 nm, $xz/yz$ Fermi wave vector difference larger than $\delta k$ = 0.014 $\pi$, and energy splitting larger than $\delta E$ = 3.5 meV. The lack of detectable ordering pinned around defects places qualitative constraints on models of fluctuating nematicity. 
\end{abstract}


\maketitle
\section{\label{sec:Intro}Introduction}

Since the 2012 discovery of enhanced high-temperature superconductivity in single-layer FeSe/SrTiO$_3$~\cite{Wang_CPL_2012}, the quest to reproduce, understand, and extend this finding remains urgent. Single-layer FeSe weakly coupled to bilayer graphene is non-superconducting down to 2.2 K~\cite{Song_PRB_2011}, but when deposited on SrTiO$_3$(001), exhibits a superconducting transition temperature $T_c$ up to 65 K~\cite{He_NatMat_2013, Tan_NatMat_2013, Lee_Nat_2014, Zhang_SciBull_2015} or 109 K~\cite{Ge_NatMat_2014}. Efforts to elucidate the microscopic mechanisms behind this transformation have presently led to divergent viewpoints~\cite{Lee_CPB_2015, Mazin_NatMat_2015}. At the crux of the debate is whether single-layer FeSe/SrTiO$_3$ exemplifies a novel pairing mechanism involving cross-interface phonon coupling, or whether it shares a common electronic mechanism with other iron chalcogenides already seen.

Indications of the first viewpoint were brought forth by angle-resolved photoemission spectroscopy (ARPES) measurements, which revealed that the primary electronic bands possess faint ``shake-off'' bands offset by 100 meV~\cite{Lee_Nat_2014, Peng_NatComm_2014}. The replication of primary band features without momentum offset suggests an electron-boson coupling sharply peaked at $\vec{q}$ $\sim$ 0. The boson was initially hypothesized to be an O phonon mode and subsequently observed on bare SrTiO$_3$(001)~\cite{Wang_arXiv_2015}. Model calculations have demonstrated that phonons can enhance spin-fluctuation-mediated pairing in FeSe~\cite{Xiang_PRB_2012, Lee_Nat_2014, Chen_arXiv_2015}. Others have argued that phonons alone can account for a significant portion of the high $T_c$~\cite{Coh_NJP_2015, Rademaker_arXiv_2015}.

An alternative but possibly complementary viewpoint is that electron doping underlies the primary enhancement of $T_c$ in single-layer FeSe/SrTiO$_3$. Early experiments observed that as-grown films become superconducting only after a vacuum annealing process~\cite{He_NatMat_2013}. This procedure presumably generates interfacial O vacancies donating electron carriers~\cite{Bang_PRB_2013}. More recent experiments showed that multilayer FeSe, which does not exhibit replica bands from coupling to SrTiO$_3$ phonons, can still develop superconductivity ($T_c$ up to 48 K suggested by ARPES) when coated with K atoms~\cite{Miyata_NatMat_2015, Wen_arXiv_2015, Ye_arXiv_2015, Tang_arXiv_2015, Tang_PRB_2015, Song_arXiv_2015}. Two observations from these latter experiments are crucial. First, the dome-shaped evolution of $T_c$ with doping refocuses attention on electronic (spin/orbital) mechanisms of pairing~\cite{Mazin_NatMat_2015}. Second, the enhanced $T_c$ emerges from a parent, bulk nematic phase, characterized in multilayer FeSe by a small orthorhombic distortion~\cite{McQueen_PRL_2009} and a large splitting of the Fe $3d_{xz}$ and $3d_{yz}$ bands~\cite{Shimojima_PRB_2014, Nakayama_PRL_2014, Watson_PRB_2015, Zhang_PRB_2015, Zhang_arXiv_2015}.

Nematic order, defined more generally as broken rotational symmetry with preserved translational symmetry, is a hallmark of the parent phase of iron-based superconductors. Importantly, both spin and orbital fluctuations that are candidate pairing glues can condense into parent nematic order~\cite{Fernandes_NatPhys_2014}. Furthermore, $\vec{q}$ $\sim$ 0 nematic fluctuations that extend beyond phase boundaries can enhance $T_c$~\cite{Fernandes_SST_2012, Yamase_PRB_2013}. This mechanism operates in any pairing channel, with increased effectiveness in a 2D system~\cite{Lederer_PRL_2015}. Recent DFT calculations have shown that bulk and single-layer FeSe exhibit a propensity towards shearing, but that strong binding to cubic SrTiO$_3$ suppresses this lattice instability~\cite{Coh_NJP_2015}. It is tempting to ask whether in addition to suppressing nematic order, this binding may push the heterostructure closer to a nematic quantum critical point, with intensified fluctuations.

To investigate the possible role of nematicity, we use scanning tunneling microscopy (STM) and quasiparticle interference (QPI) imaging. By generating scattering through moderate disorder, quasiparticle attributes such as spin/orbital/valley texture, or the superconducting order parameter, are manifested in selection rules that underlie the interference patterns. STM also affords dual real- and momentum-space visualization of electronic states within nanoscale regions. Previous STM works have uncovered $C$2 electronic patterns in parent Ca(Fe$_{1-x}$Co$_x$)$_2$As$_2$~\cite{Chuang_Science_2010, Allan_NatPhys_2013}, LaOFeAs~\cite{Zhou_PRL_2011}, NaFeAs~\cite{Rosenthal_NatPhys_2014, Cai_PRL_2014}, and superconducting orthorhombic FeSe~\cite{Song_Science_2011, Kasahara_PNAS_2014}. In addition, remnant nematic signatures were detected in the nominally tetragonal phases of NaFeAs~\cite{Rosenthal_NatPhys_2014} and FeSe$_{0.4}$Te$_{0.6}$~\cite{Singh_SciAdv_2015}.\textit{These latter observations motivate our present investigation. Can local disorder or anisotropic perturbations pin nanoscale patches of otherwise-fluctuating nematicity in single-layer FeSe/SrTiO$_3$, signaling proximate nematic quantum criticality? Or is the heterostructure too far from a nematic phase boundary for fluctuations to persist and boost $T_c$?}

This paper is organized as follows: Section~\ref{sec:Expt} presents experimental details, including QPI images acquired on single-layer FeSe/SrTiO$_3$. In order to extract the low-energy orbital texture and disentangle scattering intensities involving Fe 3$d_{xz}$ and 3$d_{yz}$ bands, we develop a $T$-matrix model of multi-orbital QPI, with results shown in Sec.~\ref{sec:Model} and mathematical details given in Appendix~\ref{sec:A}. In Sec.~\ref{sec:Nano}, we sample multiple spatial regions of our film, and based on our orbital-resolved QPI model, exclude static nematicity in the form of $xz/yz$ orbital ordering. Within domains of size $\delta r^2$ = 20 nm $\times$ 20 nm, we place quantitative bounds on $xz/yz$ Fermi wave vector difference ($\delta k$ $\leq$ 0.014 $\pi$) and $xz/yz$ pocket splitting energy ($\delta E$ $\leq$ 3.5 meV). The lack of detectable ordering pinned around impurities places qualitative constraints on models of fluctuating nematicity. A discussion and summary of results is given in Sec.~\ref{sec:Summ}. Additional details on local defect structure and fitting procedures are presented in Appendices~\ref{sec:B} and~\ref{sec:C}.

\section{\label{sec:Expt}Experiment}
\begin{figure}[t]
\includegraphics[scale=1]{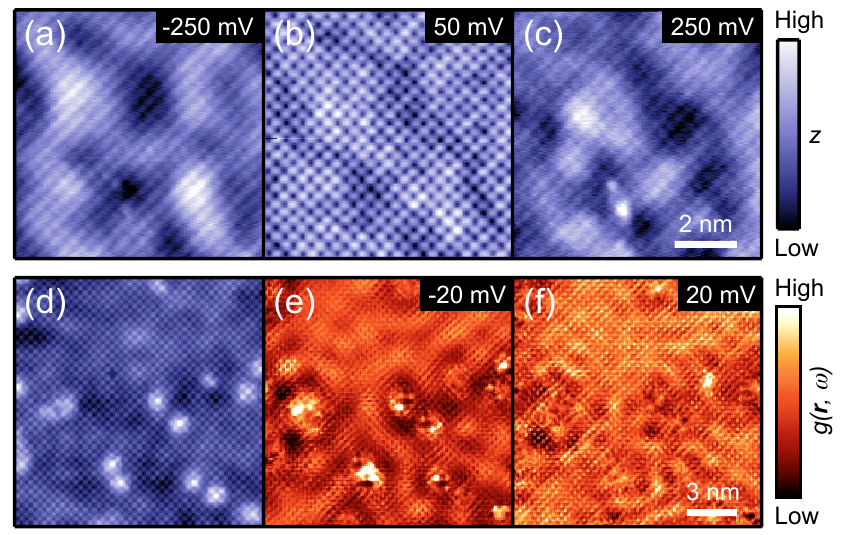}
\caption{(color online) (a)-(c) Pristine region of single-layer FeSe/SrTiO$_3$. Topographies of the same area are acquired with three set points, revealing background electronic disorder: (a) $-$250 mV, 1.25 nA, (b) 50 mV, 250 pA, (c) 250 mV, 1 nA. (d)-(f) Defect region of single-layer FeSe/SrTiO$_3$. Topography and differential tunneling conductance maps of the same area, revealing quasiparticle interference. (d) 100 mV, 5 pA, (e) $-$20 mV, 200 pA, bias oscillation $V_{\textrm{rms}}$ = 1.4 mV, (f) 20 mV, 200 pA, $V_{\textrm{rms}}$ = 1.4 mV.}
\label{Fig1}
\end{figure}

Films of single-layer FeSe were grown epitaxially on SrTiO$_3$(001) following procedures outlined in Ref.~\cite{Huang_PRL_2015}, then imaged in a home-built STM at 4.3 K. Typical superconducting gaps observed at this temperature were $\sim$14 meV~\cite{Huang_PRL_2015}. Figs.~\ref{Fig1}(a)-(c) present three atomically resolved topographies of the same area, acquired with different energy set points. Each bright spot corresponds to a surface Se atom; there are no in-plane defects in this region. Our images reveal that even pristine single-layer FeSe/SrTiO$_3$ displays appreciable electronic inhomogeneity, in strong contrast to multilayer films grown on bilayer graphene~\cite{Song_PRB_2011, Song_Science_2011}. The electronic inhomogeneity in FeSe/SrTiO$_3$ underscores the need for nanoscale measurements of electronic structure.

\begin{figure}[b]
\includegraphics[scale=1]{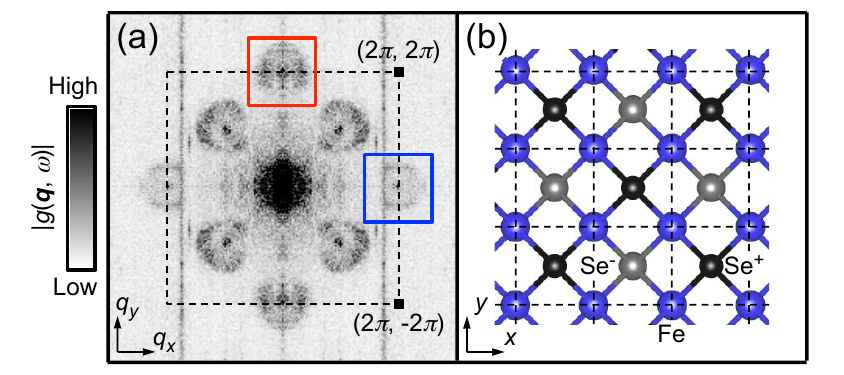}
\caption{(color online) (a) Fourier transform amplitude $|g(\vec{q}, \omega)|$ of a 20 nm $\times$ 20 nm conductance map. Set point: 20 mV, 200 pA; bias oscillation $V_{\textrm{rms}}$ = 1.4 mV. Drift-correction~\cite{Lawler_Nat_2010} and mirror symmetrization along the Fe-Fe axes are applied for increased signal. Note that fourfold rotational symmetrization has not been applied. (b) Crystal structure of single-layer FeSe. The black/gray Se atoms labeled $+$/$-$ lie above/below the plane. Dashed lines in (a) and (b) refer to the 1-Fe UC.}
\label{Fig2}
\end{figure}

To image QPI and extract local orbital information, we acquired conductance maps $g(\vec{r}, \omega)$ = $dI/dV(\vec{r}, eV)$ over regions of the film with in-plane defects (exemplified in Figs.~\ref{Fig1}(d)-\ref{Fig1}(f)). A brief commentary on the defect structures is given in Appendix~\ref{sec:B}. Figure~\ref{Fig2}(a) shows the Fourier transform amplitude $|g(\vec{q}, \omega)|$ of a map with $\omega$ = 20 meV. Ring-shaped intensites appear around $\vec{q}$ = 0, ($\pm\pi$, $\pm\pi$), (0, $\pm2\pi$), and ($\pm2\pi$, 0) due to scattering of Fermi electron pocket states. Previous works utilized ring \textit{size} dispersion to map filled- and empty-state band structure~\cite{Huang_PRL_2015}, or energy- and magnetic field-dependent ring \textit{intensities} to infer pairing symmetry from coherence factor arguments~\cite{Fan_NatPhys_2015}. Here, we will examine ring \textit{anisotropy} associated with the high-$\vec{q}$ scattering channels (red and blue boxes in Fig.~\ref{Fig2}(a)). We will demonstrate that (1) although all the QPI rings are derived from scattering within and between the same electron pockets, the high-$\vec{q}$ scattering channels have more stringent selection rules and hence a cleaner orbital interpretation; (2) the high-$\vec{q}$ scattering channels can be utilized to search for signatures of $xz/yz$ orbital ordering.

Directly from the data in Fig.~\ref{Fig2}(a), we observe an unusual relationship between the anisotropic rings around $\vec{q}$ = (0, 2$\pi$) and (2$\pi$, 0). In a single layer of FeSe, the Fe atoms are arranged in a planar square lattice, from which we define a 1-Fe unit cell (UC) [Fig.~\ref{Fig2}(b)] and crystal momentum transfer $\vec{q}$ [Fig.~\ref{Fig2}(a)]. We emphasize the distinction between the $\vec{k}$-space Brillouin zone and the $\vec{q}$-space crystal momentum transfer that is directly detected by STM imaging of QPI patterns. Including the Se atoms staggered above and below the Fe plane, the primitive UC becomes doubled. We might expect the QPI rings around (0, $2\pi$), ($2\pi$, 0) to be identical translations by the 2-Fe UC reciprocal lattice vector $2\vec{G}$ = ($-2\pi$, $2\pi$). Instead, they appear to be inequivalent and related by 90$^{\circ}$ rotation. The cause and implications of this observation will be discussed in the following section.

\section{\label{sec:Model}Multi-orbital Quasiparticle Interference}

We develop a model to map experimental $|g(\vec{q}, \omega)|$ patterns to the orbital characters of the scattered quasiparticles, similar in concept to Ref.~\cite{Zeljkovic_NatPhys_2014}. In this section, we present an intuitive picture, followed by $T$-matrix simulations with and without $xz/yz$ orbital ordering. Model details are given in Appendix~\ref{sec:A}. 

\begin{figure}[t]
\includegraphics[scale=1]{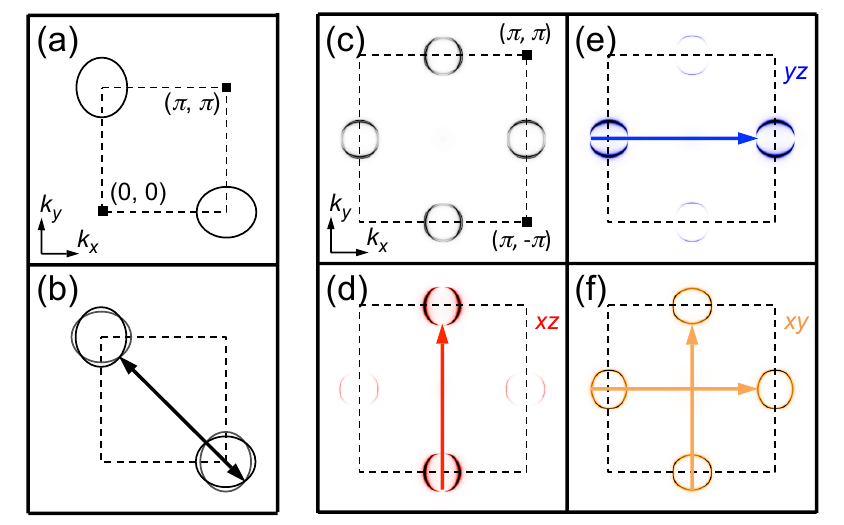}
\caption{(color online) (a) Schematic Fermi surface (FS) of single-layer FeSe/SrTiO$_3$, derived from Fe $3d$ orbitals. The FS is composed of single elliptical electron pockets around $\vec{k}$ = (0, $\pi$), ($\pi$, 0). (b) Upon introducing the potential of staggered Se atoms, the electron pockets would naively fold onto each other (double-headed arrow); however, due to incomplete folding, they remain orbitally distinct. (c) Simulated FS of single-layer FeSe/SrTiO$_3$ and (d)-(f) its dominant orbital contributions ($xz$, $yz$, $xy$). The arrows mark the expected elastic scattering wave vectors which may contribute to the $\vec{q}$ = (0, 2$\pi$) and (2$\pi$, 0) scattering channels of interest.}
\label{Fig3}
\end{figure}

Since the Fermi surface (FS) of single-layer FeSe is derived from Fe $3d$ orbitals, a natural starting point is to consider a low-energy model of a square lattice of Fe atoms. Figure~\ref{Fig3}(a) depicts a schematic FS, consisting of single elliptical electron pockets around $\vec{k}$ = (0, $\pi$), ($\pi$, 0). The hole pockets which typically appear around (0, 0) in other iron-based superconductors are sunken below the Fermi energy due to electron doping from SrTiO$_3$~\cite{Liu_NatComm_2012, He_NatMat_2013, Tan_NatMat_2013, Lee_Nat_2014}.

Although the Se atoms positioned between next-nearest neighbor Fe atoms contribute little spectral weight to the FS, their presence alters crystal symmetry and cannot be ignored. Their staggered arrangement doubles the primitive UC, folding the electron pockets around (0, $\pi$), ($\pi$, 0) on top of each other [Fig.~\ref{Fig3}(b)]. However, an underappreciated fact is that the pockets do not become identical replicas. Fe $3d$ orbitals that are even with respect to $z$-reflection ($x^2$$-$$y^2$, $xy$, $3z^2$$-$$r^2$) cannot distinguish whether Se atoms lie above/below the plane; only odd orbitals ($xz$, $yz$) feel an effective potential of doubled periodicity~\cite{Moreschini_PRL_2014}. In terms of tight-binding (TB) models, the only hopping terms that get folded in $\vec{k}$-space are those involving a product of odd and even orbitals~\cite{Lin_PRL_2011, Lin_arXiv_2014, Wang_PRL_2015}.  

To illustrate, we simulate the FS for single-layer FeSe/SrTiO$_3$ and show the dominant orbital contributions in Figs.~\ref{Fig3}(c)-\ref{Fig3}(f). Due to incomplete folding, the orbital textures of the pockets around (0, $\pi$) and ($\pi$, 0) remain distinct and separately dominated by $xz/xy$ and $yz/xy$ quasiparticles respectively. 

\begin{figure}[t]
\includegraphics[scale=1]{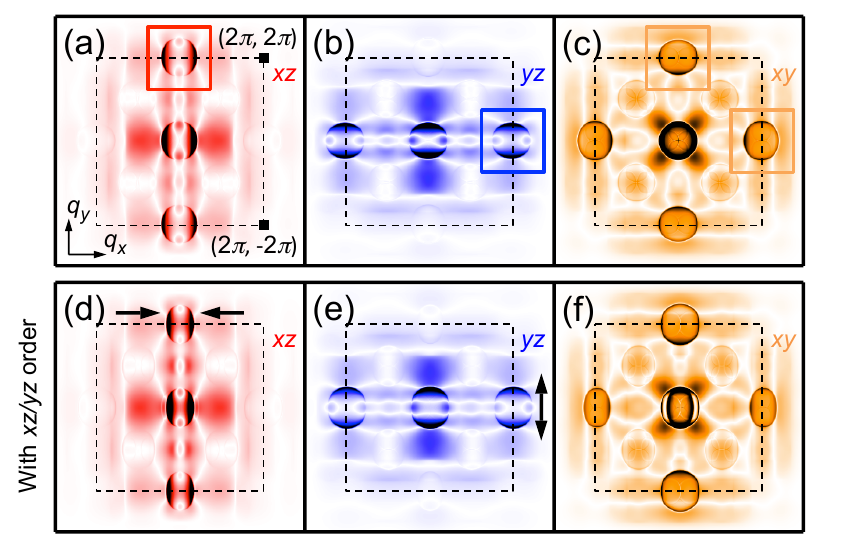}
\caption{(color online) Quasiparticle interference simulations of single-layer FeSe/SrTiO$_3$. (a)-(c) Orbital-resolved, density-of-states modulations $|\rho_{mm}(\vec{q}, \omega = 0)|$ in the presence of a localized, $s$-wave scatterer. The boxes mark signals around $\vec{q}$ = (0, $2\pi$), ($2\pi$, 0) that are the focus of this paper. (d)-(f) $|\rho_{mm}(\vec{q}, \omega = 0)|$ with the inclusion of $xz/yz$ orbital order. The arrows mark the resulting distortion of the rings.}
\label{Fig4}
\end{figure}

In the presence of disorder, elastic scattering channels should peak around wave vectors $\vec{q}$ connecting FS segments with large density of states. Considering only the pockets shown in Figs.~\ref{Fig3}(d)-\ref{Fig3}(f), we anticipate the $xz$ quasiparticles to scatter predominantly around $\vec{q}$ = (0, $2\pi$), the $yz$ quasiparticles to scatter predominantly around $\vec{q}$ = ($2\pi$, 0), and the $xy$ quasiparticles to scatter around both wave vectors. Figures~\ref{Fig4}(a)-\ref{Fig4}(c) show $T$-matrix calculations of the orbital-resolved, density-of-states (DOS) modulations $|\rho_{mm}(\vec{q}, \omega=0)|$. The index $m$ denotes the Fe $3d$ orbitals, and we assume a localized, $s$-wave scatterer in our simulations. Comparing simulation results to experimental QPI patterns at low energies [Fig.~\ref{Fig2}(a)], we observe that the elliptical rings around $\vec{q}$ = (0, $2\pi$), ($2\pi$, 0) resemble the $xz$- and $yz$-projected DOS modulations respectively. Signatures of the $xy$-projected DOS modulations, which involve oppositely oriented elliptical rings [Fig.~\ref{Fig4}(c)], appear to be suppressed in Fig.~\ref{Fig2}(a). Due to the in-plane orientation of $xy$ orbitals, their wave function amplitudes at the STM tip height are likely smaller.

The disentangling of $xz/yz$-derived QPI signals around $\vec{q}$ = (0, $2\pi$), ($2\pi$, 0) and the suppression of $xy$ signals yield a straightforward prescription to detect nanoscale $xz/yz$ orbital ordering. (In contrast, the QPI signal around $\vec{q}$ = (0, 0) would involve both $xz$ and $yz$ orbital contributions). Such orbital ordering would lead to a population imbalance of $xz/yz$ carriers, implying unequal Fermi pocket sizes and resulting anisotropy between the (0, $2\pi$), ($2\pi$, 0) scattering channels. To simulate this effect, we add on-site ferro-orbital ordering to our model. Figures~\ref{Fig4}(d)-\ref{Fig4}(f) show simulation results, which demonstrate squishing of the $xz$ ring signal around (0, $2\pi$) and the rounding of the $yz$ ring signal around ($2\pi$, 0). We show the $xy$-projected DOS modulations for completeness, although its associated tunneling amplitude is suppressed. Recent studies have also proposed orbital ordering to be bond-centered and $d$-wave~\cite{Zhang_PRB_2015, Jiang_arXiv_2015}, but these complexities produce the same qualitative effect for QPI involving the electron pockets only.

\section{\label{sec:Nano}Experimental Bounds on Nanoscale Orbital Ordering}

\begin{figure}[t]
\includegraphics[scale=1]{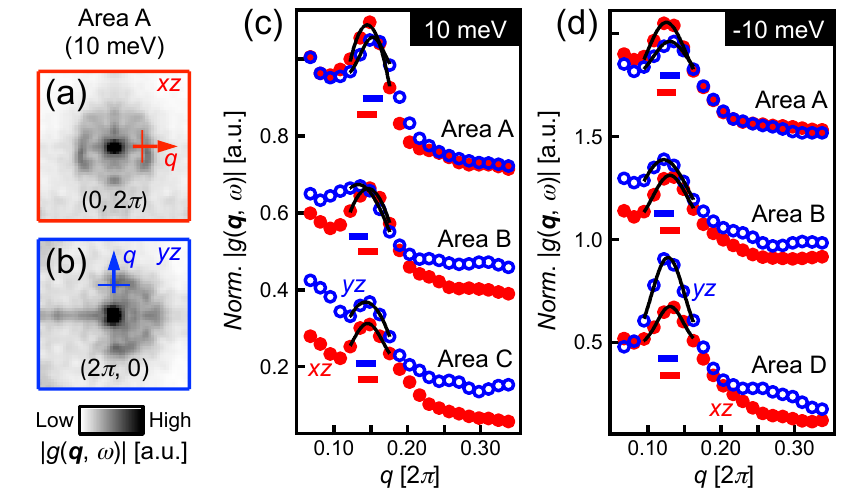}
\caption{(color online) Nanoscale wave vector analysis of orbital ordering. (a), (b) Cropped quasiparticle interference images $|g(\vec{q}, \omega)|$ around (0, $2\pi$) and ($2\pi$, 0), with arrows indicating line cuts used to compare $xz$/$yz$ Fermi pocket sizes (perpendicular bars represent averaging width). (c), (d) Line cuts from conductance maps acquired over four distinct domains ($\delta r^2$ = 20 nm $\times$ 20 nm), labeled A through D, and two energies, $\pm$10 meV. The horizontal bars mark the peak locations determined from Gaussian fits (solid lines), with inherent resolution $\delta q$ = 0.028 $\pi$. For visualization, the line cuts are normalized by the Bragg peak amplitude and vertically offset.}
\label{Fig5}
\end{figure}

We carried out experimental tests for $xz/yz$ orbital ordering as follows. To account for local inhomogeneity, we sampled QPI over four distinct domains of size $\delta r^2$ = 20 nm $\times$ 20 nm (called Areas A through D in Fig.~\ref{Fig5}). Each domain was imaged following a separate STM tip-sample approach, and was likely separated from other domains by distances larger than our scan frame width, $\delta L$ = 1.5 $\mu$m. To rule out tip anisotropy artifacts, the data from each domain were acquired with a different microscopic tip termination, modified by field emission on polycrystalline Au. Over every domain, conductance maps were acquired at low energies ($\pm$10 meV) in order to compare scattering from $xz$ and $yz$ Fermi pockets.

Figures~\ref{Fig5}(a) and \ref{Fig5}(b) show QPI images acquired over area A, cropped around $\vec{q}$ = (0, $2\pi$) and ($2\pi$, 0). We applied Gaussian smoothing with width $\sigma$ = $\delta q$, where $\delta q$ = 0.028 $\pi$ is the inherent resolution for momentum defined within a finite 20 nm $\times$ 20 nm area. To compare the $xz$- and $yz$-derived QPI rings, we took line cuts along their minor axes (arrows in Figs.~\ref{Fig5}(a) and \ref{Fig5}(b)), where the signal intensity is the strongest. Each line cut was averaged over a width of 10 pixels. The results for the four domains and two energies are shown in Figs.~\ref{Fig5}(c) and \ref{Fig5}(d). We determined peak locations from Gaussian fits (solid lines). The horizontal bars denote extracted peak locations with error $\pm \delta q$. The addition of a linear background is found to shift the fitted Gaussian peak locations, by an amount smaller than $\delta q$ (Appendix~\ref{sec:C}). In all four areas, we observe no significant deviations between the $xz$- and $yz$-derived QPI wave vectors. We therefore exclude orbital ordering with domain size larger than $\delta r^2$ = 20 nm $\times$ 20 nm and $xz/yz$ FS wave vector difference larger than $\delta k$ = $\delta q/2$ = 0.014 $\pi$ (the factor of two arises when changing between $\vec{q}$ space and $\vec{k}$ space). 

\begin{figure}[t]
\includegraphics[scale=1]{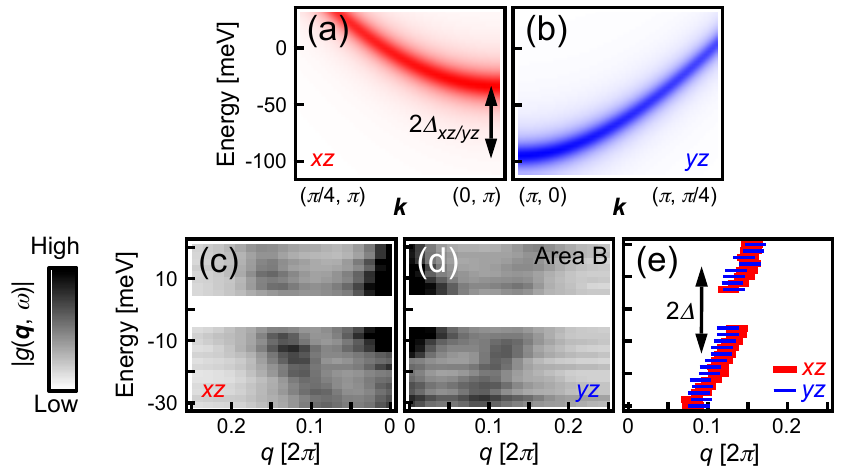}
\caption{(color online) Nanoscale dispersion analysis of orbital ordering. (a), (b) Simulated pockets with $xz/yz$ orbital ordering, revealing a split in the band edges of magnitude 2$\Delta_{xz/yz}$ = 60 meV. (c), (d) Experimental intensity plots of quasiparticle interference images $|g(\vec{q}, \omega)|$ acquired over Area B, cut along the arrows in Figs.~\ref{Fig5}(a) and \ref{Fig5}(b). (e) Plot of dispersing peaks positions extracted from Gaussian fits. The width of the horizontal bars is $\pm \delta q$ = $\pm$0.028 $\pi$. The superconducting gap magnitude is marked by 2$\Delta$.}
\label{Fig6}
\end{figure}

We also determine an energy bound on $xz/yz$ orbital ordering.  Figures~\ref{Fig6}(a) and \ref{Fig6}(b) show a simulated splitting of the $xz/yz$ bands for reference, and Figs.~\ref{Fig6}(c) and \ref{Fig6}(d) show the corresponding QPI dispersions measured over Area B. The dispersing peaks locations are extracted from Gaussian fits, shown in Fig.~\ref{Fig6}(e), and are identical within $\pm \delta q$ over the given energy range [$-$30 meV, 20 meV]. Due to the overlap with a sunken, zone center hole pocket~\cite{Liu_NatComm_2012, Huang_PRL_2015}, the lower edges of the Fermi pockets are difficult to detect. Instead, we fit the dispersing peak locations to parabolas, and find their respective band edges to be $-$51.5$\pm$3.5 meV and $-$49.6$\pm$3.0 meV. We again bound orbital ordering with $2\Delta_{xz/yz}$ $\leq$ $\delta E$ = 3.5 meV.    

\section{\label{sec:Summ}Discussion and Summary}

We return to the central debate, whether single-layer FeSe/SrTiO$_3$ exemplifies a novel interface-phonon pairing mechanism, or whether it can be explained by an electronic mechanism common to other iron chalcogenides. Recent experiments demonstrating sizeable $T_c$ in electron-doped multilayer FeSe~\cite{Miyata_NatMat_2015, Wen_arXiv_2015, Ye_arXiv_2015, Tang_arXiv_2015, Tang_PRB_2015, Song_arXiv_2015} suggest an electronic pairing mechanism in other iron chalcogenides lacking hole Fermi surfaces, distinct from a SrTiO$_3$ phonon mode. One possibility involves $\vec{q}$ $\sim$ 0 nematic fluctuations extending from the parent ordered phase. Theories have shown that such fluctuations can boost $T_c$ effectively in any pairing channel, on both the ordered and disordered sides of the phase transition~\cite{Fernandes_SST_2012, Yamase_PRB_2013, Lederer_PRL_2015}. 

Moving from multilayer to single-layer FeSe/SrTiO$_3$, we face two scenarios. One scenario is that single-layer FeSe/SrTiO$_3$ remains close to a nematic phase boundary. Here, ordering is absent, but intense nematic fluctuations may be pinned by impurities. Another scenario is that single-layer FeSe/SrTiO$_3$ lies sufficiently far away from a nematic phase boundary, such that $C$2 electronic signatures are not produced even upon local perturbation. The quantitative bounds on static $xz/yz$ orbital ordering derived from our QPI measurements favor the latter scenario. In turn, this statement would suggest that single-layer FeSe/SrTiO$_3$ is not the same as electron-doped multilayer FeSe, in which nematic fluctuations may be operative~\cite{Ye_arXiv_2015}. The addition of the SrTiO$_3$ substrate introduces novel effects beyond electron doping, such as possible interface phonons, that push the two systems apart in phase space.

To summarize, we have utilized STM and QPI imaging to demonstrate that the pronounced nematic order present in multilayer FeSe is suppressed in single-layer FeSe/SrTiO$_3$. More importantly, nanoscale nematic ordering is not recovered upon perturbation by anistropic defects. We arrived at our conclusions by comparing high-$\vec{q}$ scattering channels around (0, $2\pi$) and ($2\pi$, 0), which we showed by $T$-matrix simulations to be separately dominated by $xz$ and $yz$ quasiparticles. Our work places quantitative bounds on static $xz/yz$ orbital ordering in single-layer FeSe/SrTiO$_3$, and qualitative constraints on models of $T_c$ enhancement by nematic fluctuations.

\begin{acknowledgments}
We thank P. J. Hirschfeld, W. Ku, I. I. Mazin, and B. I. Halperin for useful conversations. We also thank A. Kreisel and S. Mukherjee for sharing an early version of their FeSe tight-binding model. This work was supported by the National Science Foundation under Grants No. DMR-0847433 and No. DMR-1231319 (STC Center for Integrated Quantum Materials), and the Gordon and Betty Moore Foundation's EPiQS Initiative through Grant No. GBMF4536. D. H. acknowledges support from an NSERC PGS-D fellowship. C. L. S. acknowledges support from the Lawrence Golub fellowship at Harvard University. S. F. and E. K. acknowledge support by Army Research Office (ARO-MURI) W911NF-14-1-0247. J. E. H. acknowledges support from the Canadian Institute for Advanced Research.
\end{acknowledgments}

\appendix

\section{\label{sec:A}Model of multi-orbital quasiparticle interference}

\textit{Model Hamiltonian:} We begin with a TB model for single-layer FeSe where the low-energy bands are projected onto the five 3$d$ orbitals of an Fe atom:
\begin{equation}\label{EqTB}
\tilde{H}^0 = \sum_{\vec{i}\vec{j}} \sum_{mn} \tilde{t}_{mn}(|i_x-j_x|,|i_y-j_y|) \tilde{c}^{\dagger}_m(\vec{i}) \tilde{c}_n(\vec{j}).
\end{equation}
Here, $\vec{i}, \vec{j}$ index the Fe lattice sites and $m, n$ index the five orbitals. The tilde symbol indicates that a momentum shift $\vec{Q}$ = ($\pi$, $\pi$) has been applied to the even orbitals in order to downfold the UC from two Fe atoms to one~\cite{Lee_PRB_2008, Lv_PRB_2011, Lin_arXiv_2014, Wang_PRL_2015}. The corresponding bare Green's function is given by
\begin{equation}\label{EqGreen}
\tilde{\vec{G}}^0(\tilde{\vec{k}}, \omega) = \big[(\omega+i \delta)\vec{I}_{5\times5} - \tilde{\vec{H}}^0(\tilde{\vec{k}})\big]^{-1},
\end{equation}
where the bolded capital symbols are matrices and $\delta$ is a broadening (= 5 meV for all simulations).

\begingroup
\begin{table*}[t]
\center
\scalebox{1}{
\setlength{\tabcolsep}{10pt}
\begin{tabular}{*9c}
\hline\hline
$t^{mn}$ & 0 & $\hat{x}$ & $\hat{y}$ & $\hat{x}+\hat{y}$ & $2\hat{x}$ & $2\hat{x}+\hat{y}$ & $\hat{x}+2\hat{y}$ & $2\hat{x}+2\hat{y}$  \\ 
\hline
$mn=11$ & -0.0192 & $-0.0538$ & $-0.1538$ & $0.0904$ & $0.0077$ & $-0.0135$ & $0.0019$ & $0.0135$\\
$mn=33$ & $-0.1538$ & $0.1051$ &  & $-0.0404$ & $-0.0077$ &  &  & \\
$mn=44$ & $0.0462$ & $0.0885$ &  & $0.0577$ & $-0.0115$ & $-0.0115$ &  & $-0.0115$\\
$mn=55$ & $-0.1504$ & $-0.0385$ &  &  & $-0.0154$ & $0.0077$ &  & $-0.0038$\\
\\
$mn=12$ &  &  &  & $0.0192$ &  & $-0.0058$ &  & $0.0135$\\
$mn=13$ &  & $-0.1362$ &  & $0.0381$ &  & $0.0081$ &  & \\
$mn=14$ &  & $0.1304$ &  & $0.0054$ &  & $0.0108$ &  & \\
$mn=15$ &  & $-0.0762$ &  & $-0.0327$ &  &  &  & $-0.0054$\\
$mn=34$ &  &  &  &  &  & $-0.0038$ &  & \\
$mn=35$ &  & $-0.1154$ &  &  &  & $-0.0077$ &  & \\
$mn=45$ &  &  &  & $-0.0577$ &  &  &  & $0.0038$\\
\hline\hline
\end{tabular}}
\caption{Rescaled hopping parameters for tight-binding model adapted from Ref.~\cite{Graser_NJP_2009}. Here, $m$=1 is $xz$, $m$=2 is $yz$, $m$=3 is $x^2$--$y^2$, $m$=4 is $xy$, $m$=5 is $3z^2$--$r^2$.}
\label{Table2}
\end{table*}
\endgroup

\begin{figure}[b]
\includegraphics[scale=1]{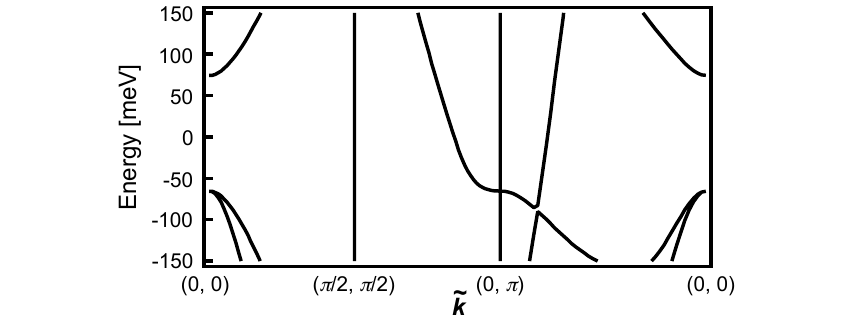}
\caption{Band structure of single-layer FeSe/SrTiO$_3$ in the 1-Fe Brillouin zone. The tilde symbol indicates that a momentum shift $\vec{Q}$ = ($\pi$, $\pi$) has been applied to the even orbitals to downfold the unit cell from two Fe atoms to one. Adapted from Ref.~\cite{Graser_NJP_2009}, with rescaled hopping parameters to match the pocket edges measured in Ref.~\cite{Huang_PRL_2015}.}
\label{Fig7}
\end{figure}

We adapt hopping parameters $\tilde{t}_{mn}$ computed in Ref.~\cite{Graser_NJP_2009}, then apply rescaling to qualitatively capture the low-energy spectrum of single-layer FeSe/SrTiO$_3$~\cite{Huang_PRL_2015}. The hopping terms are given in Table~\ref{Table2}, and the resulting band structure is shown in Fig.~\ref{Fig7}. The electron pocket around $\tilde{\vec{k}}$ = (0, $\pi$) remains attached to a hole pocket, but this does not affect our simulation results closer to the Fermi level. The positions of the $\Gamma$ pockets above and below the Fermi energy also do not affect our simulations.

\textit{Fermi surface simulations:} Computing the FS of single-layer FeSe/SrTiO$_3$ [Figs.~\ref{Fig3}(c)-\ref{Fig3}(f)] from our TB model requires that we restore the original crystal symmetry (with a 2-Fe UC) induced by staggered Se atoms. To do so, we transform the lattice operators with a site-dependent sign factor~\cite{Lin_arXiv_2014}:
\begin{equation}\label{trans1}
c^{\dagger}_m(\vec{i}) = (-p_m)^{-i_x-i_y} \tilde{c}^{\dagger}_m(\vec{i}).
\end{equation}
Here, $p_m$ = $\pm$1 for orbitals that are even/odd with respect to $z$-reflection. This transformation is equivalent to undoing the downfolding operation applied in Eq.~\ref{EqTB} and shifting the even orbitals by $-\vec{Q}$ in momentum space:
\begin{equation}\label{trans}
   c^{\dagger}_m(\vec{k}) = 
     \begin{cases}
       \tilde{c}^{\dagger}_m(\vec{k}), & p_m = -1, \\
       \tilde{c}^{\dagger}_m(\vec{k}_{\vec{Q}}), & p_m = +1,
     \end{cases}
\end{equation}
where $\vec{k}_{\vec{Q}} = \vec{k} - \vec{Q}$. The orbital components of the FS are given by
\begin{equation}\label{EqFS}
   A^0_{mm}(\vec{k}, \omega = 0) = 
     \begin{cases}
       \tilde{A}^0_{mm}(\vec{k}, 0), & p_m = -1, \\
       \tilde{A}^0_{mm}(\vec{k}_{\vec{Q}}, 0), & p_m = +1,
     \end{cases}
\end{equation}
where $\tilde{A}^0_{mm}(\tilde{\vec{k}}, \omega)$ = $-$Im $\tilde{G}^0_{mm}(\tilde{\vec{k}}, \omega)$/$\pi$. Further insights on (1) the connection between Eq.~\ref{EqFS} and ARPES-measured band structures, (2) common misconceptions of whether spectroscopic probes measure quasiparticles closer to the 1-Fe or 2-Fe Brillouin zone description, and (3) proper folding of the superconducting pairing structure, are given in Refs.~\cite{Lin_PRL_2011, Lin_arXiv_2014, Wang_PRL_2015}.

\textit{Quasiparticle intereference simulations:} To generate QPI, we introduce a localized, $s$-wave scatterer at $\vec{i}$ = (0, 0) of uniform strength $V$ = 1 eV in all orbital channels. The resulting impurity Green's function is given by
\begin{equation}\label{EqimpGreen}
\tilde{\vec{G}}(\tilde{\vec{k}}, \tilde{\vec{k}}', \omega) = \tilde{\vec{G}}^0(\tilde{\vec{k}}, \omega) \vec{T}(\omega) \tilde{\vec{G}}^0(\tilde{\vec{k}}', \omega), 
\end{equation}
for $\tilde{\vec{k}}$ $\neq$ $\tilde{\vec{k}}'$, and the $T$-matrix is momentum-independent:
\begin{equation}\label{EqT}
\vec{T}(\omega) = \bigg[\vec{I}_{5\times5}-V\int\frac{d^2\tilde{k}}{(2\pi)^2} \tilde{\vec{G}}^0(\tilde{\vec{k}}, \omega) \bigg]^{-1} V.
\end{equation}

\begin{figure}[t]
\includegraphics[scale=1]{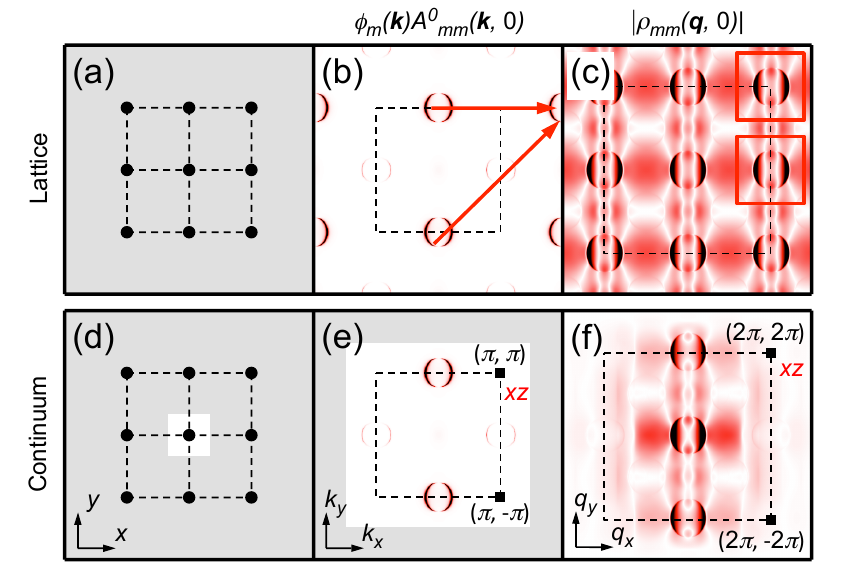}
\caption{(color online) (a) Lattice model, in which the Green's function has nonzero weight restricted to discrete lattice points; i.e., the Wannier functions are given by $\phi(\vec{r}-\vec{i}) = \delta(\vec{r}-\vec{i})$. (b) In momentum space, $\phi_m(\vec{k}) = 1$, such that there is no cutoff for states involved in scattering. Shown here is $\phi_m(\vec{k})A^{0}_{mm} (\vec{k}, \omega = 0)$ for the $xz$ orbital. (c) Consequently, additional ring intensities arise in $|\rho_{mm}(\vec{q}, 0)|$ around $\vec{q} = (2\pi, 0), (2\pi, 2\pi)$ (arrows in (b), boxes in (c)) that are not observed experimentally. (d) Continuum model, which incorporates nonlocal effects due to a finite Wannier function width (white square). (e) We model the experimental data with Wannier functions of the form $\phi_m(\vec{k})$ = 1 for $k_x$, $k_y$ $\in$ [-1.5$\pi$, 1.5$\pi$], and $\phi_m(\vec{k})$ = 0 otherwise (white square). (f) As a result, there are fewer scattering channels.}
\label{Fig8}
\end{figure}

Since STM measures local density of states in real space, we additionally transform lattice operators $\tilde{c}^{\dagger}_m(\vec{i})$ into continuum operators $\psi^{\dagger}_m(\vec{r})$:
\begin{equation}\label{Eqpsi}
\psi^{\dagger}_m(\vec{r}) = \sum_{\vec{i}} (-p_m)^{-i_x-i_y} \phi_m^*(\vec{r}-\vec{i}) \tilde{c}^{\dagger}_m(\vec{i}).
\end{equation}
The first factor on the right recovers the proper crystal symmetry (2-Fe UC) due to staggered Se atoms [Eq.~\ref{trans1}]. The second factor on the right, $\phi_m$, is the Wannier function associated with orbital $m$ at site $\vec{i}$. This factor captures nonlocal tunneling contributions~\cite{Kreisel_PRL_2015}. For simplicity, we approximate the Wannier functions at the STM tip height with a square cutoff in momentum space: $\phi_m(\vec{k})$ = 1 for $k_x$, $k_y$ $\in$ [-1.5$\pi$, 1.5$\pi$], and $\phi_m(\vec{k})$ = 0 otherwise. In real space, this corresponds to a characteristic tunneling width of 0.67 ($a_{\textrm{Fe-Fe}}$), which is neeeded to reproduce experimental QPI patterns. Figure~\ref{Fig8} illustrates qualitative differences between simulations with (continuum model) and without (lattice model) non-local tunneling.

From Eq.~\ref{Eqpsi}, we obtain the continuum impurity Green's function:
\begin{multline}\label{EqcontGreen}
\mathcal{G}_{mm}(\vec{k}, \vec{k'}, \omega) \\ =
	\begin{cases}
	\tilde{G}_{mm}(\vec{k}, \vec{k'}, \omega) \phi^*_m(\vec{k}) \phi_m(\vec{k'}), & p_m = -1, \\
        \tilde{G}_{mm}(\vec{k}_{\vec{Q}}, \vec{k'}_{\vec{Q}}, \omega) \phi^*_m(\vec{k}) \phi_m(\vec{k'}), & p_m = +1.
     \end{cases}
\end{multline}
Only diagonal elements are shown for brevity. Finally, the orbital projections of the DOS modulations are given by
\begin{multline}
\rho_{mm}(\vec{q}, \omega) = \frac{i}{2\pi} \int \frac{d^2k}{(2\pi)^2} \bigg[\mathcal{G}_{mm}(\vec{k}, \vec{k}+\vec{q}, \omega) \\
-\mathcal{G}^*_{mm}(\vec{k}, \vec{k}-\vec{q}, \omega)\bigg].
\end{multline}
Figs.~\ref{Fig4}(a)-(c) show plots of $|\rho_{mm}(\vec{q}, 0)|$ for the $xz$, $yz$, and $xy$ orbitals.

\textit{Orbital ordering:} To simulate on-site, ferro-orbital ordering, we include the following term~\cite{Mukherjee_PRL_2015} in our TB Hamiltonian [Eq.~\ref{EqTB}]:
\begin{equation}\label{EqOO}
\tilde{H}^0_{xz/yz} = \Delta_{xz/yz} \sum_{\vec{i}} \big[\tilde{c}^{\dagger}_{xz}(\vec{i}) \tilde{c}_{xz}(\vec{i}) - \tilde{c}^{\dagger}_{yz}(\vec{i}) \tilde{c}_{yz}(\vec{i})\big].
\end{equation}
A value of $\Delta_{xz/yz}$ = 30 meV was used for Figs.~\ref{Fig4}(d)-(f) and \ref{Fig6}(a)-(b). 

\textit{Superconductivity:} The inclusion of superconductivity does not change the QPI orbital texture. Following Ref.~\cite{Chi_PRB_2014}, we introduce superconductivity in band space, but compute scattering in orbital space. From the normal-state TB Hamiltonian [Eq.~\ref{EqTB}], we define bands $\tilde{\vec{\epsilon}}(\tilde{\vec{k}}) = \tilde{\vec{U}}(\tilde{\vec{k}})\tilde{\vec{H}}^0(\tilde{\vec{k}})\tilde{\vec{U}}^{\dagger}(\tilde{\vec{k}})$, where $\tilde{\vec{U}}(\tilde{\vec{k}})$ represents a unitary transformation. The Green's function in the superconducting state is then given by 
\begin{equation}
\tilde{\vec{G}}^0_{\textrm{BCS}}(\tilde{\vec{k}}, \omega) = \big[(\omega+i \delta)\vec{I}_{10\times10} - \tilde{\vec{H}}^0_{\textrm{BCS}}(\tilde{\vec{k}})\big]^{-1},
\end{equation}
where $\tilde{\vec{H}}^0_{\textrm{BCS}}(\tilde{\vec{k}})$ has the following form in Nambu representation:
\begin{multline}
\tilde{\vec{H}}^0_{\textrm{BCS}}(\tilde{\vec{k}}) = \\
\begin{pmatrix}
\tilde{\vec{U}}^{\dagger}(\tilde{\vec{k}})\tilde{\vec{\epsilon}}(\tilde{\vec{k}})\tilde{\vec{U}}(\tilde{\vec{k}}) & \tilde{\vec{U}}^{\dagger}(\tilde{\vec{k}})\vec{\Delta}(\tilde{\vec{k}})\tilde{\vec{U}}^*(-\tilde{\vec{k}}) \\
\tilde{\vec{U}}^{T}(-\tilde{\vec{k}})\vec{\Delta}^*(\tilde{\vec{k}})\tilde{\vec{U}}(\tilde{\vec{k}}) & -\tilde{\vec{U}}^{T}(-\tilde{\vec{k}})\tilde{\vec{\epsilon}}(-\tilde{\vec{k}})\tilde{\vec{U}}^*(-\tilde{\vec{k}})
\end{pmatrix}.
\end{multline}
We model isotropic gaps in band space: $\vec{\Delta}(\tilde{\vec{k}}) = \Delta \vec{I}_{5\times5}$, with $\Delta$ = 14 meV based on our $dI/dV$ measurements. (Recent ARPES measurements have detected small gap anisotropy~\cite{Zhang_arXiv_2015(2)}). We also take the impurity potential of a localized, non-magnetic, $s$-wave scatterer:
\begin{equation}
\vec{V} = 
\begin{pmatrix}
V\vec{I}_{5\times5} & 0 \\
0 & -V\vec{I}_{5\times5} \\
\end{pmatrix}.
\end{equation}

\begin{figure}[t]
\includegraphics[scale=1]{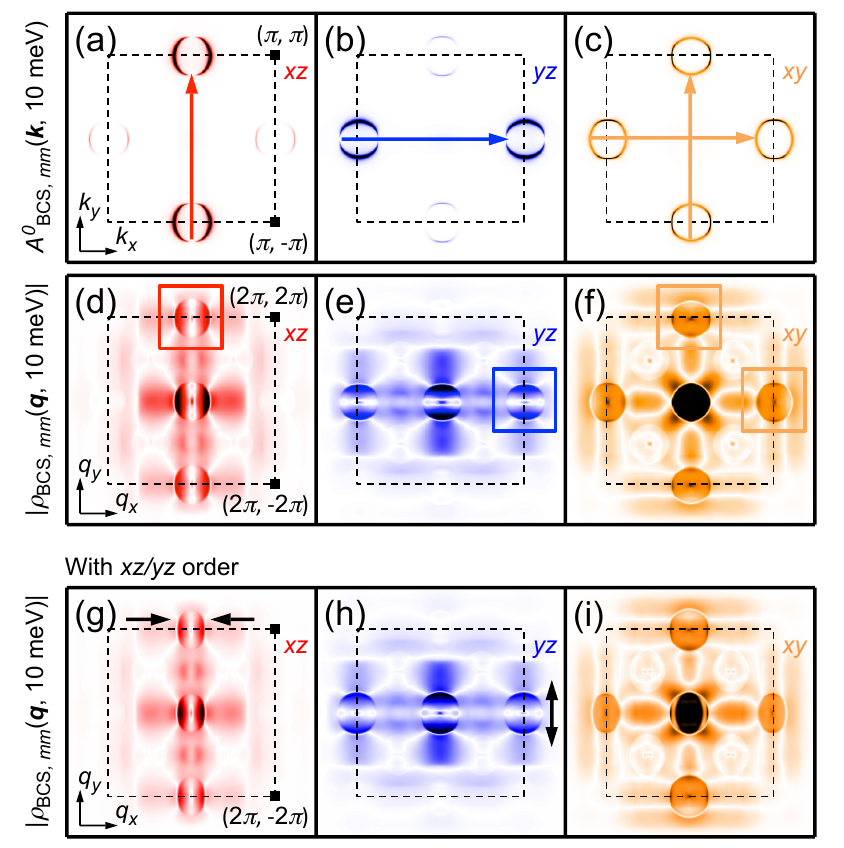}
\caption{(color online) Quasiparticle interference simulations in the superconducting state, with isotropic gaps of 14 meV on all bands. (a)-(c) Orbital-resolved spectral function $A^0_{\textrm{BCS}, mm}(\vec{q}, 10~\textrm{meV})$. (d)-(f) Orbital-resolved density-of-states modulations $|\rho_{\textrm{BCS}, mm}(\vec{q}, 10~\textrm{meV})|$ in the presence of a localized, non-magnetic, $s$-wave scatterer. (g)-(i) Same as (d)-(f), but including $xz/yz$ orbital ordering ($\Delta_{xz/yz}$ = 30 meV).}
\label{Fig9}
\end{figure}

Figure~\ref{Fig9} shows QPI simulations with the inclusion of superconductivity, at energy $\omega$ = 10 meV. There is little difference compared with the normal-state calculations, without or with $xz/yz$ orbital ordering.

\begin{figure}[b]
\includegraphics[scale=1]{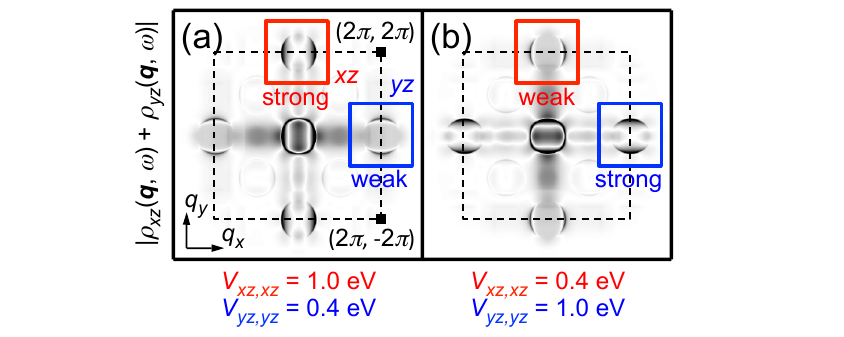}
\caption{QPI simulations in the presence of a localized, anisotropic scatterer. While the relative intensities of the $xz/yz$ scattering channels around $\vec{q}$ = (0, 2$\pi$) and (2$\pi$, 0) are modified, their wave vectors remain unchanged.}
\label{Fig10}
\end{figure}

\textit{Anisotropic scatterer:} In an angular momentum expansion of the $T$-matrix, the leading component should be $s$-wave; i.e., intraorbital scattering, with $V_{mn} = V_{mm}\delta_{mn}$, should dominate. $V_{mm}$ can in general vary with orbital, but this simply modifies the relative weights of the orbital-resolved DOS modulations. In Fig.~\ref{Fig10}, we illustrate this effect in the case of a $C$2 scatterer ($V_{xz, xz} \neq V_{yz, yz}$). Tuning the strengths of $V_{xz, xz}$ and $V_{yz, yz}$ tunes the intensity of the $\vec{q}$ = (0, 2$\pi$) and ($2\pi$, 0) scattering channels respectively; however, the scattering wave vectors remain unchanged and are a more robust measure of orbital ordering. On the other hand, the $\vec{q}$ $\sim$ (0, 0) channel will display anisotropies related to the scattering potential, so we do not analyze it.

\textit{Equivalent ten-orbital formulation:} We derive an equivalent formulation of the continuum impurity Green's function for ten-orbital TB models, such as those in Refs.~\cite{Eschrig_PRB_2009, Mukherjee_PRL_2015}. The ten $3d$ orbitals come from the two inequivalent Fe atoms ($A$ and $B$) of the primitive UC: $(xy)^A$, $(x^2-y^2)^A$, $(ixz)^A$, $(iyz)^A$, $(z^2)^A$, $(xy)^B$, $(x^2-y^2)^B$, $(-ixz)^B$, $(-iyz)^B$, $(z^2)^B$.

We begin with the real-space representation of the continuum impurity Green's function [Eq.~\ref{EqcontGreen}] for a five-orbital TB model: 
\begin{multline}\label{EqrsGreen}
\mathcal{G}_{mn}(\vec{r}, \vec{r'}, \omega) = \sum_{\vec{i}\vec{j}} (-p_m)^{-i_x-i_y} (-p_n)^{j_x+j_y}\\
\phi^*_m({\vec{r}-\vec{i}}) \phi_n({\vec{r'}-\vec{j}}) \tilde{G}_{mn}(\vec{i}, \vec{j}, \omega). 
\end{multline}
Here, $\tilde{G}_{mn}(\vec{i}, \vec{j}, \omega)$ is the lattice impurity Green's function, given in momentum space by Eq.~\ref{EqimpGreen}. Next, we decompose the Fe lattice into two sublattices:
\begin{equation}\label{Eqlattices}
\begin{cases}
A = \{ \vec{i}: i_x + i_y = \textrm{odd} \}, \\
B = \{ \vec{i}: i_x + i_y = \textrm{even} \}.
\end{cases}
\end{equation}
For diagonal terms ($m = n$), the sum in Eq.~\ref{EqrsGreen} can be split into four contributions:
\begin{multline}
\label{Eqsplit}
\sum_{\vec{i}, \vec{j}} \big(. . .\big) (-p_m)^{-i_x-i_y+j_x+j_y} = \sum_{\vec{i} \in A, \vec{j} \in A} \big(. . .\big) (+1)  \\
+ \sum_{\vec{i} \in A, \vec{j} \in B} \big(. . .\big) (-p_m) + \sum_{\vec{i} \in B, \vec{j} \in A} \big(. . .\big) (-p_m) \\ 
+ \sum_{\vec{i} \in B, \vec{j} \in B} \big(. . .\big) (+1).
\end{multline}
Taking the Fourier transform of Eq.~\ref{Eqsplit} yields
\begin{multline}\label{Eqten}
\mathcal{G}_{mm}(\vec{k}, \vec{k'}, \omega) = \big[ \tilde{\mathbb{G}}_{m,m}(\vec{k}, \vec{k'}, \omega) \\ + p_m \tilde{\mathbb{G}}_{m+5,m}(\vec{k}, \vec{k'}, \omega) + p_m \tilde{\mathbb{G}}_{m,m+5}(\vec{k}, \vec{k'}, \omega) \\ + \tilde{\mathbb{G}}_{m+5,m+5}(\vec{k}, \vec{k'}, \omega) \big] \phi^*_m(\vec{k}) \phi_m(\vec{k'}).
\end{multline}
Here, $m$ runs from 1 through 5, $\mathcal{G}_{mm}(\vec{k}, \vec{k'}, \omega)$ is derived from a five-orbital TB model, and $\tilde{\mathbb{G}}_{mn}(\vec{k}, \vec{k'}, \omega)$ is the lattice Green's function for a ten-orbital TB model. The $p_m$ factors appear because of minus signs present in the orbital definitions of $(-ixz)^B$, $(-iyz)^B$. The middle terms in Eq.~\ref{Eqten}, which mix orbitals $m$ and $m+5$, represent intraorbital basis site interference. Importantly, the sum of these terms are non-zero for a finite Wannier function width. These crucial terms, which have not been considered in previous ten-orbital QPI calculations of Fe-SCs~\cite{Chi_PRB_2014}, are required in order to reconcile five-orbital and ten-orbital QPI calculations in the presence of non-local tunneling.

\begin{figure}[t]
\includegraphics[scale=1]{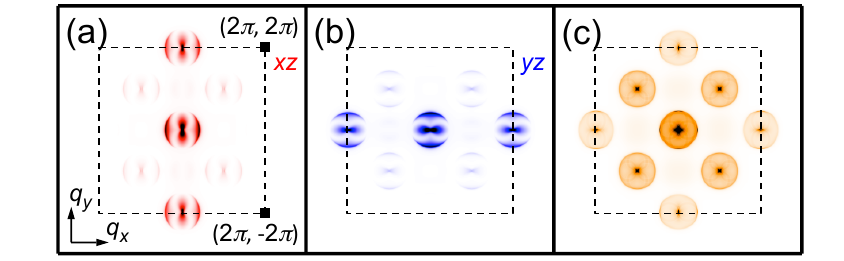}
\caption{Quasiparticle interference simulations derived from calculating the joint density of states separately for each orbital.}
\label{Fig11}
\end{figure}

\textit{Joint density of states:} Figure~\ref{Fig11} demonstrates that our $T$-matrix simulations can be qualitatively approximated by calculating the joint DOS separately for each orbital:
\begin{multline}
\rho_{mm}(\vec{q}, \omega) \sim \\ 
\int d^2 k A^0_{mm}(\vec{k}, \omega) A^0_{mm}(\vec{k} + \vec{q}, \omega) \phi^*_m(\vec{k}) \phi_m(\vec{k} + \vec{q}).
\end{multline}

\section{\label{sec:B}Symmetry breaking in local defect structures}

\begin{figure}[t]
\includegraphics[scale=1]{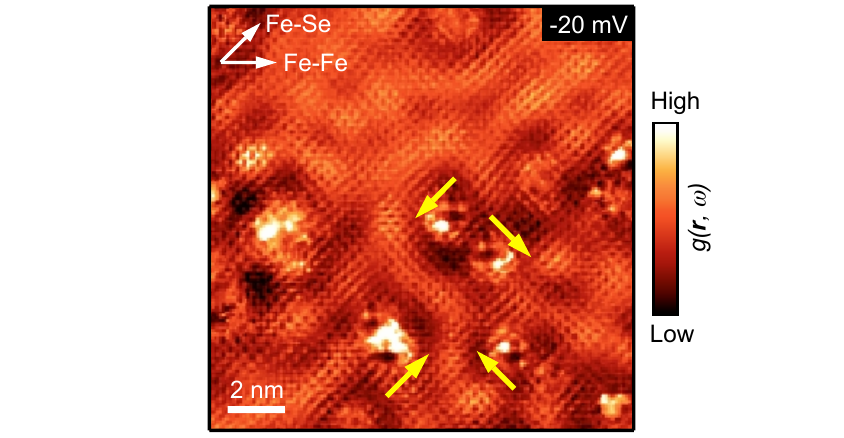}
\caption{(color online) Differential tunneling conductance map revealing dominant type of anisotropic impurities (reproduced from Fig.~\ref{Fig1}(e)). The defects are directed along the crystalline Fe-Se axes and appear in four possible orientations (yellow arrows). Set point: $-$20 mV, 200 pA, bias oscillation $V_{\textrm{rms}}$ = 1.4 mV.}
\label{Fig12}
\end{figure}

In NaFeAs~\cite{Rosenthal_NatPhys_2014, Cai_PRL_2014}, multilayer FeSe~\cite{Song_Science_2011}, and Ca(Fe$_{1−x}$Co$_x$)$_2$As$_2$~\cite{Allan_NatPhys_2013}, the dominant atomic-scale defects have been shown to pin larger electronic dimers that are unidirectional within nanoscale domains and aligned along one Fe-Fe axis, providing evidence of local nematic ordering. Here we search for a similar effect in single-layer FeSe/SrTiO$_3$.

A closer inspection of Fig.~\ref{Fig1}(e) reveals a dominant in-plane defect that appears as adjacent bright and dark atoms along the crystalline Fe-Se axes [Fig.~\ref{Fig12}], and exists along all four orientations, like impurities observed in LiFeAs~\cite{Grothe_PRB_2012, Hanaguri_PRB_2012}. These defects are similar in their atomic-scale structure and Fe-Se orientation to the dominant defects in NaFeAs~\cite{Rosenthal_NatPhys_2014, Cai_PRL_2014} and multilayer FeSe~\cite{Song_Science_2011}. However, the defects observed in single-layer FeSe/SrTiO$_3$ do not show the larger Fe-Fe electronic dimers. Furthermore, in Sec.~\ref{sec:Nano} we considered four different nanoscale domains (20 nm $\times$ 20 nm), each containing several such randomly-oriented defects, but our nanoscale wave vector and dispersion analyses found no significant difference in electronic response between the Fe-Fe axes. The chance that the impurity orientations would exactly balance in all four sampled regions is small.

\section{\label{sec:C}Fitting details} 

Here we detail the fitting procedures used to derive wave vector bounds on nanoscale orbital ordering [Figs.~\ref{Fig5} and~\ref{Fig6}]. We performed Gaussian fits using the iterative Levenberg-Marquardt algorithm implemented by Igor Pro. Fit errors were estimated from residuals and represent one standard deviation of the fit coefficient from its true value, assuming independent and identically-distributed Gaussian noise. As seen in Table~\ref{Table1}, the fit errors ($\delta q_0$) for $q_0$ are insignificant compared to the data resolution $\delta q$. Note that $\sigma_q$ gives the correlation length of the QPI patterns, rather than the uncertainty in its wavevector. 

\begin{figure}[b]
\includegraphics[scale=1]{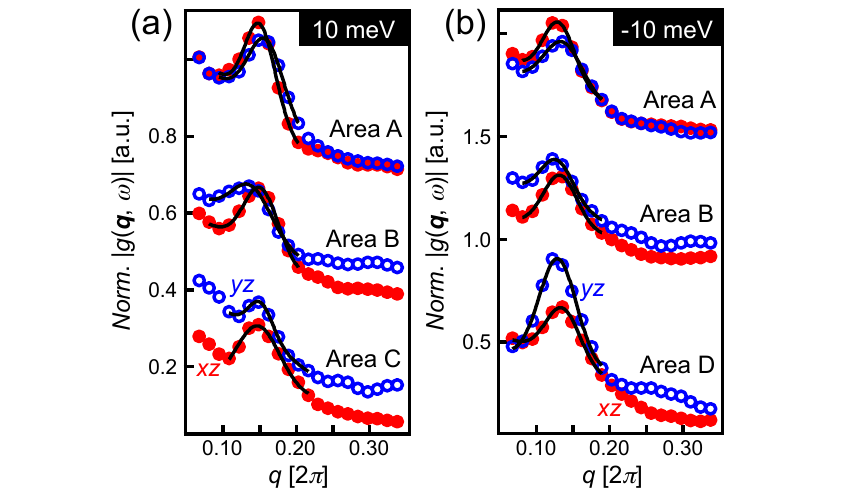}
\caption{(color online) Line cuts from conductance maps, reproduced from Figs.~\ref{Fig5}(c)-(d). The solid lines denote Gaussian fits with linear background. Fit parameters are recorded in Table~\ref{Table1}.}
\label{Fig13}
\end{figure}

Since there is a sizeable slope in the line cuts, we also performed Gaussian fitting with linear background for comparison [Fig.~\ref{Fig13} and Table~\ref{Table1}]. We find that the $q_0$ values are shifted by amounts smaller than the data resolution $\delta q$. The leading error is therefore the data resolution $\delta q$, which we report as short horizontal blue and red bars in Figs.~\ref{Fig5}(c) and (d). 

\begingroup
\begin{table*}[t]
\center
\scalebox{1}{
\setlength{\tabcolsep}{10pt}
\begin{tabular}{ c c c c c r }
  \hline
     Line cut label & \multicolumn{2}{c}{\underline{Gaussian plus constant background}} & \multicolumn{2}{c}{\underline{Gaussian plus linear background}} \\
     & $q_0 \pm \delta q_0$ [2$\pi$] & $\sigma_{q} \pm \delta \sigma_{q}$ [2$\pi$] & $q_0 \pm \delta q_0$ [2$\pi$] & $\sigma_{q} \pm \delta \sigma_{q}$ [2$\pi$]\\
  \hline\hline
  Area A, 10 meV, $xz$ & 0.145$\pm$0.002 & 0.08$\pm$0.70 & 0.1506$\pm$0.0007 & 0.021$\pm$0.01 \\
  Area A, 10 meV, $yz$ & 0.152$\pm$0.001 & 0.017$\pm$0.004 & 0.1584$\pm$0.0004 & 0.0238$\pm$0.0006 \\
  Area B, 10 meV, $xz$ & 0.146$\pm$0.001 & 0.08$\pm$0.60 & 0.1518$\pm$0.0006 & 0.0233$\pm$0.0009 \\
  Area B, 10 meV, $yz$ & 0.133$\pm$0.002 & 0.05$\pm$0.04 & 0.139$\pm$0.002 & 0.027$\pm$0.003 \\
  Area C, 10 meV, $xz$ & 0.1456$\pm$0.0005 & 0.021$\pm$0.004 & 0.148$\pm$0.003 & 0.029$\pm$0.004 \\
  Area C, 10 meV, $yz$ & 0.1429$\pm$0.0009 & 0.1$\pm$0.5 & 0.152$\pm$0.001 & 0.017$\pm$0.001 \\
  \hline
  Area A, $-$10 meV, $xz$ & 0.125$\pm$0.001 & 0.026$\pm$0.009 & 0.1307$\pm$0.0008 & 0.021$\pm$0.001 \\
  Area A, $-$10 meV, $yz$ & 0.130$\pm$0.002 & 0.02$\pm$0.01 & 0.1363$\pm$0.0004 & 0.0257$\pm$0.006 \\
  Area B, $-$10 meV, $xz$ & 0.1300$\pm$0.0003 & 0.023$\pm$0.002 & 0.1324$\pm$0.0006 & 0.0228$\pm$0.0008 \\
  Area B, $-$10 meV, $yz$ & 0.1217$\pm$0.0007 & 0.033$\pm$0.007 & 0.1281$\pm$0.0009 & 0.022$\pm$0.001 \\
  Area D, $-$10 meV, $xz$ & 0.130$\pm$0.001 & 0.018$\pm$0.003 & 0.1361$\pm$0.0009 & 0.023$\pm$0.001 \\
  Area D, $-$10 meV, $yz$ & 0.1275$\pm$0.0006 & 0.022$\pm$0.004 & 0.129$\pm$0.001 & 0.025$\pm$0.002 \\
  \hline
\end{tabular}}
\caption{Comparison of fit parameters between (1) Gaussians with constant background, $g = g_0 + A\exp (-(q-q_0)^2/(2\sigma_{q}^2))$ [Figs.~\ref{Fig5}(c)-(d)], and (2) Gaussians with linear background, $g = g_0 + mx + A\exp (-(q-q_0)^2/(2\sigma_{q}^2))$ [Fig.~\ref{Fig13}]. As reference, the data resolution $\delta q$ = 0.014 $2\pi$.}
\label{Table1}
\end{table*}
\endgroup

%

\end{document}